\documentstyle[psfig]{mn}

\title{A complete infrared Einstein ring in the gravitational lens system B1938+666}
\author[L.J. King et al.]{L.J. King$^{1}$, N. Jackson$^{1}$,\\
\mbox{} \\
{\LARGE R.D. Blandford$^{2}$, M.N. Bremer$^{3}$, I.W.A. Browne$^{1}$, A.G. de Bruyn$^{4,5}$,}\\
\mbox{}\\
{\LARGE C. Fassnacht$^{2}$, L. Koopmans$^{4}$,  D. Marlow$^{1}$, S. Nair$^{1}$, P.N. Wilkinson}\\
\mbox{}\\
1. University of Manchester, NRAL Jodrell Bank, Macclesfield, 
Cheshire SK11 9DL, England\\
2. California Institute of Technology, Pasadena, CA 91125, USA\\
3. Institut d'Astrophysique de Paris, 98bis Blvd Arago, 75014 Paris, France\\
4. Netherlands Foundation for Radio Astronomy, Postbus 2, 7990AA
Dwingeloo, Netherlands\\
5. Kapteyn Astronomical Institute, Postbus 800, 9700AV Groningen,
Netherlands\\
}
\date{14th October 1997}
\onecolumn
\begin{document}
\baselineskip 18pt
\maketitle
\vskip 3mm
\begin{abstract}
We report the discovery, using NICMOS on the Hubble Space Telescope, of
an arcsecond-diameter Einstein ring in the gravitational lens system
B1938+666. The lensing galaxy is also detected, and is most likely 
an early-type. Modelling of the ring is presented and compared with the
radio structure from MERLIN maps. We show that the Einstein ring is
consistent with the gravitational lensing of an extended infrared component, 
centred between the two radio components. 
\end{abstract}

\section{Introduction}
In 1992 Patnaik et al. (1992a) published the first results of the JVAS (Jodrell
Bank--VLA Astrometric Survey). The primary purpose of the JVAS was to survey
flat-spectrum radio sources from the 5-GHz NRAO catalogues (Condon \&
Broderick 1985; Condon \& Broderick 1986; Condon, Broderick \& Seielstad
1989), in order to identify phase reference sources for MERLIN. Six 
gravitational lens systems have been found in this survey: B0218+357
(Patnaik et al. 1993), MG0414+0534 (this object was
originally discovered in the MIT-Greenbank survey; Hewitt et al. 1992), 
B1422+231 (Patnaik et al. 1992b), B1030+074 (Xanthopoulos et al. 1997,
in preparation), B2114+022 (Augusto et al. 1997, in preparation), 
and B1938+666 (Patnaik et al. 1992a; King et al. 1997), the subject 
of this paper.

B1938+666 has a complex radio structure, with a dominant arc and two 
pairs of fairly compact components contained within an arcsecond 
separation (Patnaik et al. 1992a; King et
al. 1997). King et al. (1997) provide a simple model of the system 
in which the lensed object consists of three components. The first
component (R2) lies in the 3-image region of the source plane, is
compact, has a steep spectrum, and produces a double radio image 
separated by about 1 arcsecond. The second (R1) and third components 
both lie in the 5-image region, producing four detectable images. (The
third component is a weak component close to R1 and will not be included
in the model presented here). Two images, one of R1 and one of the third
component, combine to form a relatively compact condensation at the 
southern end of the system. 

Rhoads, Malhotra \& Kundi{\'c} (1996) reported infrared and optical imaging 
observations of B1938+666, taken at the Apache Point Observatory. They 
detected an object coincident with the radio position of the lens system,
with a very red optical--infra-red colour 
$(r (0.65\mu{\rm m}) - K' (2.12\mu{\rm m}) = 6.8 \pm 0.25)$. The angular 
resolution of
 their observations was not high enough to determine whether the object 
detected was the lensing galaxy or the lensed source. 

\section{HST observations and results}

Observations of B1938+666 were obtained on 1997 August 13 with the Near
Infrared Camera/Multi-Object Spectrograph (NICMOS) attached to the
Hubble Space Telescope. Four observations, of 3840s, 1088s, 3840s and
2048s were obtained, giving a total integration time of just over three
hours. Two orbits of spacecraft time were used during a Continuous
Viewing Zone opportunity, allowing continuous observations of the target
throughout the spacecraft orbit.

The NIC-1 camera was used for these observations, which has a pixel
scale of 43 mas and a resolution of approximately the diffraction limit
of 0\farcs 14. The F160W filter was used, which has a peak wavelength of
1.583$\mu$m and a FWHM of 0.40$\mu$m, and corresponds approximately to
the standard infrared H-band. All observations were calibrated by the 
standard STScI pipeline procedure  involving bias
and dark current subtraction, linearity and flat-field correction,
photometric calibration and cosmic ray identification and removal. All
observations were taken using MULTIACCUM mode, which consists of
multiple non-destructive readouts of the detectors. This allows good
cosmic ray detection and removal. The
final image has been formed by coaddition of the four individual
exposures and rotated using the ORIENTAT keyword in the file
header.

Only one, quite complex, object is visible in the $\sim$11-arcsecond field of
view. This object consists of a central condensation, surrounded by a ring
of diameter 0\farcs 95. More extended fuzz is also visible out to a radius
of $\sim2^{\prime\prime}$. We identify the central condensation as the
centre of the lensing galaxy, and the extended fuzz as its outer
regions. The lensing galaxy has a compact, bright core and low
ellipticity, resembles HST images of E-type galaxies (e.g. Driver et al.
1995) and is thus likely to be an early-type elliptical.
The lensed object is extended in the optical, and some part of
it is lensed into the ring structure that we see in the HST picture. We
present and discuss more detailed models of the system below.

The total H magnitude of all components visible is measured as
18.0$\pm$0.3, with an error dominated by the relatively bright and
variable background. Because of the rough agreement with the H magnitude
obtained by Rhoads et al. (1996), 17.6$\pm$0.1, and because no other
object is visible in our field, it is likely that the object seen in the
HST image is the same as the Rhoads et al. (1996) identification. The
1\farcs6 -- 1\farcs8 seeing obtained by Rhoads et al., however, would
not have been sufficient to resolve the ring emission, which contributes
$\sim$40\% to the total H-band flux.

\section{Lens model and discussion}

In this section we show that the Einstein ring and ring-peaks in the NICMOS 
picture of B1938+666 are consistent with the gravitational lensing of an 
infrared source close to the radio sources. This is achieved though lens
modelling, and comparison of the infrared and radio structures. 

The lens models were obtained using the ${\cal AIPS}$ task GLENS, which 
models the action of a gravitational lens on a background source. We do 
not use a more complex lens modelling algorithm in this paper since our 
primary aim is to show that the appearance of B1938+666 on the NICMOS picture
can be attributed to lensing, and also because the source and lens redshifts
remain undetermined. In GLENS, the lens was represented by an isolated 
elliptical potential well of the asymptotically isothermal form discussed by 
Blandford \& Kochanek (1987), whose projected centre in the source 
plane was specified by a pixel position. Unlensed source components were 
represented by circular Gaussians and were also assigned pixel positions in 
the source plane. The position and magnification of images formed when a 
source is lensed are related to the first and second derivatives of the 
lensing potential respectively; see for example Blandford \& Kochanek (1987). 

\begin{figure}
\centering
\setlength{\unitlength}{1in}
\begin{picture}(3.5,4)
\put(0,0){\includegraphics{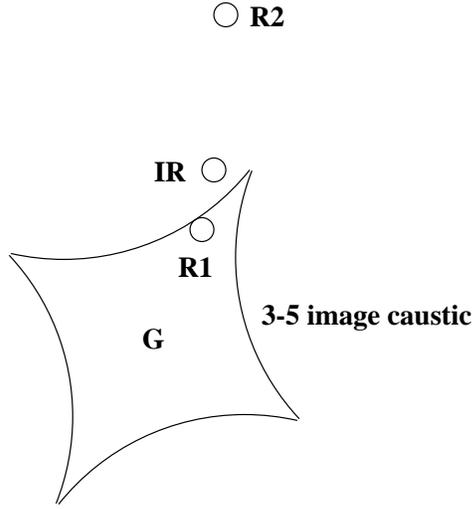}}
\end{picture}
\caption{A schematic for the source plane of B1938+666, with circles 
indicating the intrinsic positions of the radio components R1 and R2 
and the infrared component IR, with respect to the 3-5 image caustic. 
The lensing galaxy centre is marked G, and is 51 mas from the radio
component R1. The infra-red component IR is
extended all the way into the caustic; the parts which lie close to the
middle of the caustic produce the ring. Component R1 produces the arc
complex at the northern end of the radio structure, together with the
most southerly compact component; R2 is doubly imaged, producing the two
compact components outside the ring. See Fig. 6 of King et al. (1997)
for a more detailed description of how the radio components map on to
the image plane.\label{caus}}
\end{figure}

We used GLENS to obtain a simple lens model for the 5-GHz MERLIN map
of 1938+666, using the model and multi-frequency observations
presented in King et al. (1997) as a starting point. The MERLIN map
was made using the Caltech package DIFMAP procedure (Shepherd
1997). The source plane was divided into a 512$\times$512 grid, with
each pixel corresponding to 3.1~mas at the source redshift. For the
lensing galaxy the ellipticity, $\epsilon$, was fixed at a modest
value of 0.07, which is consistent with the Einstein ring and the
appearance of the lensing galaxy itself on the NICMOS picture. The
position angle of the major axis of the galaxy (N through E) was fixed at
150$^{\tiny o}$, which results in a good radio (and NICMOS) image
model. The main features on the radio map were reproduced by
representing the unlensed source by two Gaussian components and
iteratively varying the components' positions, running GLENS, and
comparing the image plane output with the 5-GHz map. We convolved the
image plane output for our best model to a resolution similar to the
5-GHz MERLIN map (which has a beam of 47$\times$44 mas in PA=$-9^{\circ}$)
using the ${\cal AIPS}$ task CONVL. Figure \ref{caus} indicates the 
positions of the primary radio
components R1 and R2 with respect to the 3-5 image caustic.

To obtain a model for the NICMOS picture we used the same parameters to 
describe the lensing galaxy as in the radio model. What do we know about 
the infra-red source that would be lensed to reproduce the features of the NICMOS
picture? There are two main peaks in the ring on the NICMOS picture, 
consistent with a source centred in the 3-image region of the source 
plane, close to radio component R2. If the infra-red emission was centred 
in the 5-image region coincident with R1, there would be a strong 
peak at a position angle similar to that of the arc-complex observed 
in the radio map. That there is a complete ring in the NICMOS picture suggests
 that the source of the infra-red emission is much larger than that of the radio 
emission. Our initial model for the unlensed source therefore consisted of a 
Gaussian centred on the position of R2, covering the 3-5 image caustic. 

\begin{figure}
\centering
\setlength{\unitlength}{1in}
\begin{picture}(3.5,4)
\put(0,0){\includegraphics{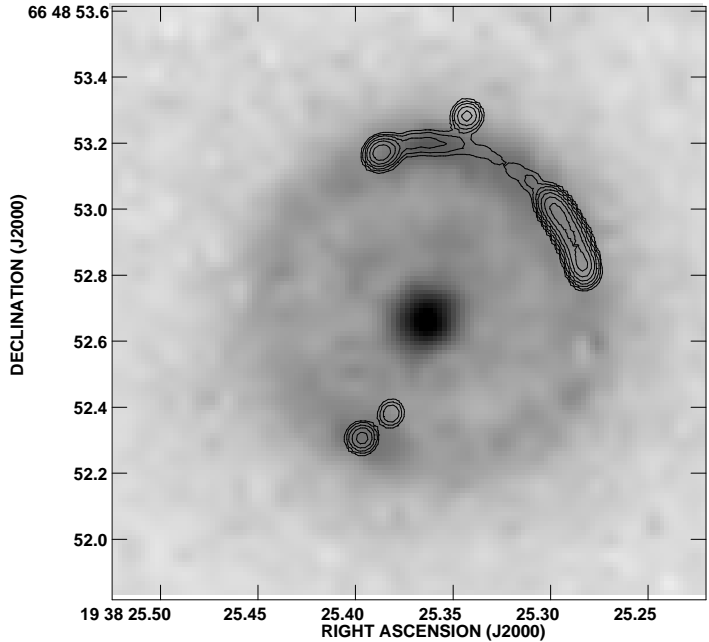}}
\end{picture}
\caption{The 5-GHz MERLIN map of B1938+666 (logarithmic contours with the
lowest contour at 0.5 mJy/beam), and the NICMOS picture as a greyscale.
\label{data}}
\end{figure}

The positional accuracy of NICMOS is about 0.5-arcsec, so we 
cannot directly identify corresponding positions in the radio map and NICMOS 
picture, although the radio arc must straddle the critical line
which corresponds to the image of the caustic separating the 3-5 image regions, 
and also traces the position of the Einstein ring. To correct for the
discrepancy between the radio and optical reference frames, the radio model 
and map were superimposed and J2000 coordinates corresponding to 
the pixel position of the lens were determined. Since the position of the 
lens was the same in the radio and optical models and the lens is visible on 
the NICMOS picture, it can be used to fix the optical coordinate reference 
frame. In Figure \ref{data} we show the radio map as contours with the NICMOS 
picture as a greyscale. 

\begin{figure}
\centering
\setlength{\unitlength}{1in}
\begin{picture}(3.5,4)
\put(0,0){\includegraphics{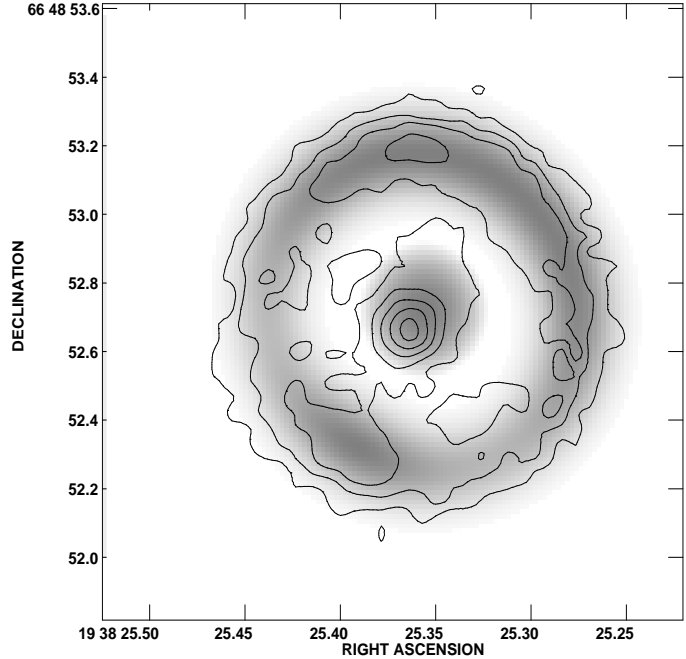}}
\end{picture}
\caption{The NICMOS picture of B1938+666 (contoured at 
0.01 $\times$ 2,2.5,3,4,5,6,7,8 counts/s) and the GLENS NICMOS image model 
as a greyscale. Note that the central blob of the image model
corresponds to the position of the lensed object in the absence of
lensing, whereas the central
bright object in the NICMOS picture corresponds to the lensing galaxy. 
The lensing galaxy positions in the image and model agree well.\label{ir}}
\end{figure}

When the initial NICMOS image model was put on the same coordinate
frame as the NICMOS picture, the peaks in the model and picture had a
small offset. We corrected for this by moving the model infra-red source
slightly closer to the 3-5 image caustic (marked IR on Figure
\ref{caus}); Figure \ref{ir} shows the NICMOS picture (contours) and
image model (greyscale). This produces good agreement between the
model and the observed NICMOS images. It should be noted that the 
greyscale image includes a representation of the unlensed source at its
centre. The NICMOS picture itself reveals the lensing galaxy slightly
displaced from the position of the unlensed source, as expected. The
relative positions of the radio components and the infrared source as
required by the model, as shown in Figure \ref{caus}, indicate
that the background object is a compact radio galaxy; the radio lobes
are disposed symmetrically either side of the host galaxy.
 
About 60\% of the total emission observed in the infrared is contributed by the
lensing galaxy. The lensed object is likely to be magnified by a factor of
about 10--20 (see e.g. the formula of Refsdal \& Surdej 1995), giving an
intrinsic magnitude for the lensed object of $\sim$21 in H. This is
likely to correspond to K$\sim$22, fainter by about 3 magnitudes than
typical brightnesses of radio galaxies at z=2 (Dunlop et al. 1989). 
This low magnitude may be a consequence of reddening in the lensing
galaxy. This could be tested by HST observations using WFPC2 to
determine the optical--infrared colours of the lensed source and lensing
galaxy. The lensing galaxy has an infra-red magnitude similar to that of the
galaxies of Dunlop et al. at z$\sim$1-2. Further high sensitivity 
radio observations may reveal a radio core coincident 
with the infrared counterpart.

\section{Summary}

We have discovered an infrared Einstein ring in the lens system
B1938+666, using NICMOS on the HST. The lensing galaxy is also
detected well above the noise. We have compared the infrared and radio
structures, and demonstrated that the ring is consistent with the
gravitational lensing of an extended component, probably a galaxy,
centred between the two main radio components. More detailed modelling
and HST imaging in other wavebands will pin down the properties of
the lensing galaxy and the intrinsic positions of the radio and
infrared components.

\section*{Acknowledgments}
 
This research used observations with the Hubble Space Telescope, 
obtained at the Space Telescope Science Institute, which is operated 
by Associated Universities for Research in Astronomy Inc. under NASA 
contract NAS5-26555. 
MERLIN is operated as a National Facility by NRAL, University of Manchester,
on behalf of the UK Particle Physics \& Astronomy Research Council. 
This research was supported by European Commission, TMR Programme, Research 
Network Contract ERBFMRXCT96-0034 ``CERES". We thank Mark Lacy and 
Hanadi AbdelSalam for useful discussions. LJK is grateful to the 
Astrophysics Department,
University of Oxford, for hospitality during part of the writing of this
paper.

\end{document}